\documentclass[fleqn,10pt]{SelfArx} 
\usepackage[english]{babel} 
\usepackage{pifont}

\PassOptionsToPackage{hyphens}{url}
\usepackage{hyperref} 

\definecolor{color1}{RGB}{0,0,90} 
\definecolor{color2}{RGB}{0,20,20} 

\hypersetup{
	hidelinks,
	colorlinks,
	breaklinks=true,
	urlcolor=color2,
	citecolor=color1,
	linkcolor=color1,
	bookmarksopen=false,
	pdftitle={Title},
	pdfauthor={Author},
}

\JournalInfo{Extended Technical Report, 2020} 
\Archive{Vol. I} 

\PaperTitle{Characterising attacks targeting low-cost routers: a MikroTik case study (Extended)}

\Authors{João M. Ceron\textsuperscript{1}*, Christian Scholten\textsuperscript{2}, Aiko Pras\textsuperscript{2}, Elmer Lastdrager\textsuperscript{1}, Jair Santanna\textsuperscript{3} } 

\affiliation{\textsuperscript{1}\textit{SIDN Labs, The Netherlands}}
\affiliation{\textsuperscript{2}\textit{University of Twente, The Netherlands}}
\affiliation{\textsuperscript{3}\textit{Northwave Security, The Nevertheless}}
\affiliation{*\textbf{Corresponding author}: joao.ceron@sidn.nl} 

\newcommand{\keywordname}{Keywords} 

\Abstract{Attacks targeting network infrastructure devices
pose a threat to the security of the internet. An attack targeting
such devices can affect an entire autonomous system. In recent years,
malware such VPNFilter, Navidade and SonarDNS has been used to
compromise low-cost routers and commit all sorts of cybercrimes from
DDoS attacks to ransomware deployments. Routers of the type concerned
are used both to provide last-mile access for home users and to manage
interdomain routing (BGP). MikroTik is a particular brand of low-cost
router. In our previous research, we found more than 4 million
MikroTik routers available on the internet. We have shown that these
devices are also popular in Internet Exchange infrastructures. Despite
their popularity, these devices are known to have numerous
vulnerabilities. In this paper, we extend our previous analysis by
presenting a long-term investigation of MikroTik-targeted attacks. By
using a highly interactive honeypot that we developed, we collected
more than 44 million packets over 120 days, from sensors deployed in
Australia, Brazil, China, India, the Netherlands and the United
States. The incoming traffic was classified on the basis of Common
Vulnerabilities and Exposures to detect attacks targeting MikroTik
devices. That enabled us to identify a wide range of activities on the
system, such as cryptocurrency mining, DNS server redirection and more
than 3,000 successfully established tunnels used for eavesdropping.
Although this research focus on Mikrotik devices, both the methodology
and the publicly available scripts can be easily applied to any other
type of network device.
}

\begin{document}

\maketitle 

\tableofcontents 

\thispagestyle{empty} 


\section{Introduction}

Network infrastructure devices have been actively exploited by cyber
actors~\cite{US-CERT:TA18-106A,US-CERT:TA16-250A}. A variety of
attacks can be carried out by abusing such devices. In 2018, more than
half a million low-cost routers were infected by the VPNFilter
malware~\cite{norton2018misc}. With a view to disrupting that malware
campaign, the Federal Bureau of Investigation
(FBI)~\cite{fbi2018misc} issued an urgent request for users to reboot
their routers. In the same year, there were several other campaigns
aimed at low-cost routers (e.g.  GhostDNS malware, Navidade and
SonarDNS)~\cite{zdnet2019misc}. Infrastructure devices can be used
for last-mile access as well as to manage interdomain routing (BGP).

Half of the core routers used in one of the biggest internet exchanges
in the world (connecting 1467 autonomous
systems)~\cite{certbr2019misc} are manufactured by MikroTik. This
manufacturer uses the same operating system (RouterOS) for all their
low-cost routers (used by home users and in the core network
infrastructure). In order to improve the security of this type of
router and set proper defences, it is crucial to understand the risk
and characteristics of the attacks that target these routers.

One effective way to investigate attacks targeting devices connected
to the internet is to use a honeypot. As demonstrated by Lobato
\textit{et al.}~\cite{lobato2018adaptive}, honeypots are valuable
resources for detecting new or unknown attacks targeting a system. The
network management and operations community has been using honeypots
and similar approaches to improve insight into malicious activities
within
networks~\cite{docarmo2011im,nassar2007im,franccois2011bottrack,sperotto2010overview}.
In this paper, we extend our previous analysis by presenting a
long-term investigation of MikroTik-targeted attacks. Although we
focus our analysis on MikroTik routers, any other low-cost router
could have been used. The main contributions made in this paper are as
follows.

\begin{itemize}

    \item \textbf{We shed light on the global landscape of MikroTik
    devices.} To that end, we investigate (1) how many MikroTik
    devices are reachable via the internet, (2) what port numbers are
    most commonly open on such devices, and (3) where such devices are
    located. This contribution highlights the importance of
    investigating MikroTik devices and facilitates definition of a
    realistic MikroTik device honeypot.

    \item \textbf{We propose a realistic, easily deployed honeypot that mimics low-cost MikroTik routers.}
    The proposed honeypot uses virtualisation to run the system in the
    cloud and enables remote management and security implementation
    mechanisms by using a set of modules. Our honeypot image is publicly
    available on the project website~\cite{mikrotik2019exposed}. The
    study's first contribution (MikroTik device landscape) enables us
    to define and deploy a realistic honeypot by indicating (1) where
    to place the sensors, (2) which port numbers should be open for
    interaction with attackers, and (3) what ethical issues the design
    needs to address.

   \item \textbf{We propose an automated classification of the traffic
    collected in the honeypot and discuss ways of mitigating attacks of
    the types identified}. The classification facilitates the
    quantification of the attacks and is based on two databases of
    manually defined attack signatures. The signatures developed for this
    study are publicly available on the project website and are compatible
    with the Berkeley Packet Filter. Finally, we discuss how to mitigate
    the attacks on MikroTik devices.

\end{itemize}

We compiled the landscape of MikroTik devices using more than 4 TB of
data (from Shodan.io). For the classification of attacks, we used more
than 44 million packets and 3.8 million log records.

The remainder of this paper is structured as follows. First, in
\autoref{sec:relatedwork}, we describe the uniqueness of this work in
relation to the state of the art. Then, in  \autoref{sec:landscape},
we describe the methodology and the results of our survey of the
global MikroTik device landscape.  In \autoref{sec:honeypotdesign}, we
describe the design, implementation and limitations of our MikroTik
device honeypot. In \autoref{sec:results}, we first discuss how we
automatically classify the traffic collected in the honeypot. Then we
present and discuss our findings based on 120 days of collected data.
Finally, in \autoref{sec:conclusion}, we discuss our conclusions and
the directions that future research is likely to take.

\section{Related Work}
\label{sec:relatedwork}

Honeypots are usually designed to facilitate the improvement of system
security by enabling the study of malicious behaviours. Do Carmo
\textit{et al.} \cite{docarmo2011im} and Nassar \textit{et al.}
\cite{nassar2007im}, for example, described the use of honeypots for
improving the security of VoIP systems. There are other works that,
although they do not propose a specific honeypot, set out strategies
for detecting malicious behavior by performing network measurements.
For example, François \textit{et al.} \cite{franccois2011bottrack}
analysed communication behavior patterns to infer potential botnet
activities; and Sperotto \textit{et al.} \cite{sperotto2010overview}
showed how flow-based techniques can be used to detect scans, worms,
botnets and denial-of-service attacks.  Our work is similar to the
work described by those authors. However, we focus our attention on
low-cost devices that are (also) used in the core infrastructures of
networks. Such devices play a crucial role in the networks of
developing countries (as described in the next section).

Various papers have also been published analysing vulnerabilities in
generic low-cost routers and proposing countermeasures. For example,
Niemietz and Schwenk \cite{ROUTERSEC:RUB:2015}  evaluated home
routers and showed how these routers are vulnerable to cross-site
scripting attacks and User-Interface (UI) redressing; Mujtaba and
Nanda \cite{BGPANALYSIS:EDCOUN:2011}  analysed vulnerabilities in the
BGP protocol on low-cost routers; Ghourabi \textit{et al.}
\cite{HONEYPOT:IEETR:2009} discussed the abuse of another routing
protocol used by low-cost routers, the RIP protocol. While the papers
in question focus on generic low- cost routers, only Mazdadi
\textit{et al.} \cite{ROUTEROSFORENSICS:IJCSIS:2017} investigated
MikroTik devices.  However, the research in question was limited to
the analysis of a single attack and did not propose a reproducible way
of monitoring and characterising attacks.

Of the previous studies we found, the one most similar to our work is
by Baines~\cite{baines2019misc}. He provides insights regarding the
global landscape of MikroTik devices and their vulnerabilities. The
main difference between his work and ours is that all his observations
are based on active scanning of port number 8291. By contrast, we
begin by analysing the ports that MikroTik devices actually use in
practice.  Then we use the findings to create a realistic MikroTik
honeypot. We believe that our approach provides a more complete
understanding of the landscape and the prevalence of vulnerabilities.
We have also classified more attacks than are described in Baines'
\cite{baines2019misc} work.

Finally, and crucially for our work, most investigations of honeypots
involve discussion of the ethical and legal perspectives of using
honeypots for research. Sokol and
Andrejko\cite{HONEYPOTSLIABILITY:SPRINGER:2015} discussed the issue
of liability in relation to honeypot use. The question of liability
can arise if honeypots are exploited by attackers and used to launch
attacks. The paper in question discusses the systems that need to be
taken into account when designing a honeypot, in order to minimise the
risks. Similarly, Hecker \textit{et al.} \cite{HONEYPOT:MARY:2006}
argued for the use of dynamic honeypots instead of low or
high-interaction honeypots. We took all the lessons described in those
papers into account in the design of our honeypot, with a view to
minimising issues.

\section{MikroTik Devices Landscape}
\label{sec:landscape}

For investigating the global landscape of MikroTik devices, we rely on
the information collected by the \url{Shodan.io} project. This project
port-scans the entire IPv4 address space. Although other similar
projects could be used, \textit{e.g.} \url{Censys.io}, Shodan covers
more generic ports (giving a higher chance of port-scanning the
particular ports used by MikroTik devices). In order to reduce the
amount of traffic generated, Shodan scans only a set of IP addresses
and a set of service ports each day. It takes around two weeks to scan
the entire IPv4 space. Shodan also tries to retrieve the responses
from services running at the relevant IP addresses (\textit{i.e.} the
banners), which are used to, for example, classify the types of
service running on the devices (\textit{i.e.} the products).

Initially, we entered several different queries on Shodan’s online
platform: “mikrotik” (1,657,859 results), “product: mikrotik”
(3,700,193 results) and “product:mikrotiksmb” (1,323 results). Each
result is related to an IP address and an open service port. It means
that each product/service port running at an IP address is one single
entry in Shodan’s dataset. In order to determine the precise number of
devices involved, we would need to merge all the results relating to
MikroTik. After noticing that we were unable to download all results
(more than 1M) for further analysis, we contacted Shodan's operators,
who granted us access to one month of their dataset.

\vspace{0.3cm}
\textbf{Dataset and methodology.} Each day of Shodan’s scanning yields
a file of around 130 GB\@. We downloaded the data from the period 17
July 2019 to 17 August 2019. In total, more than 4 TB of data was
retrieved. We filtered the entire dataset for records that contained
the string ‘mikrotik’. Hence, we know that, in every entry in the
resulting data subset, the banner (response from a device) definitely
relates to a MikroTik device (true positive).

\vspace{0.3cm}

\textbf{Limitations.} First, if a banner from a MikroTik device is
empty, neither Shodan nor we are able to classify it, possibly
implying in false negative. Second, Shodan does not scan all possible
service ports, implying that devices will not be found if they use a
particular unscanned port. Third, although Shodan updates their
dataset every two weeks (accordingly to the owner of the project), we
sought to determine the total number of devices over a one-month
period, without flushing the data. The implication is that we did not
remove IP addresses that may have ceased to point to MikroTik devices.
Finally, we expect that a small number of devices are not discoverable
by Shodan because they are, for example, behind NAT systems, or do not
answer on any open service port, or are not online at the time of
measurement.

\vspace{0.3cm}

\begin{figure}[ht]
    \centering
	\includegraphics[width=0.8\linewidth]{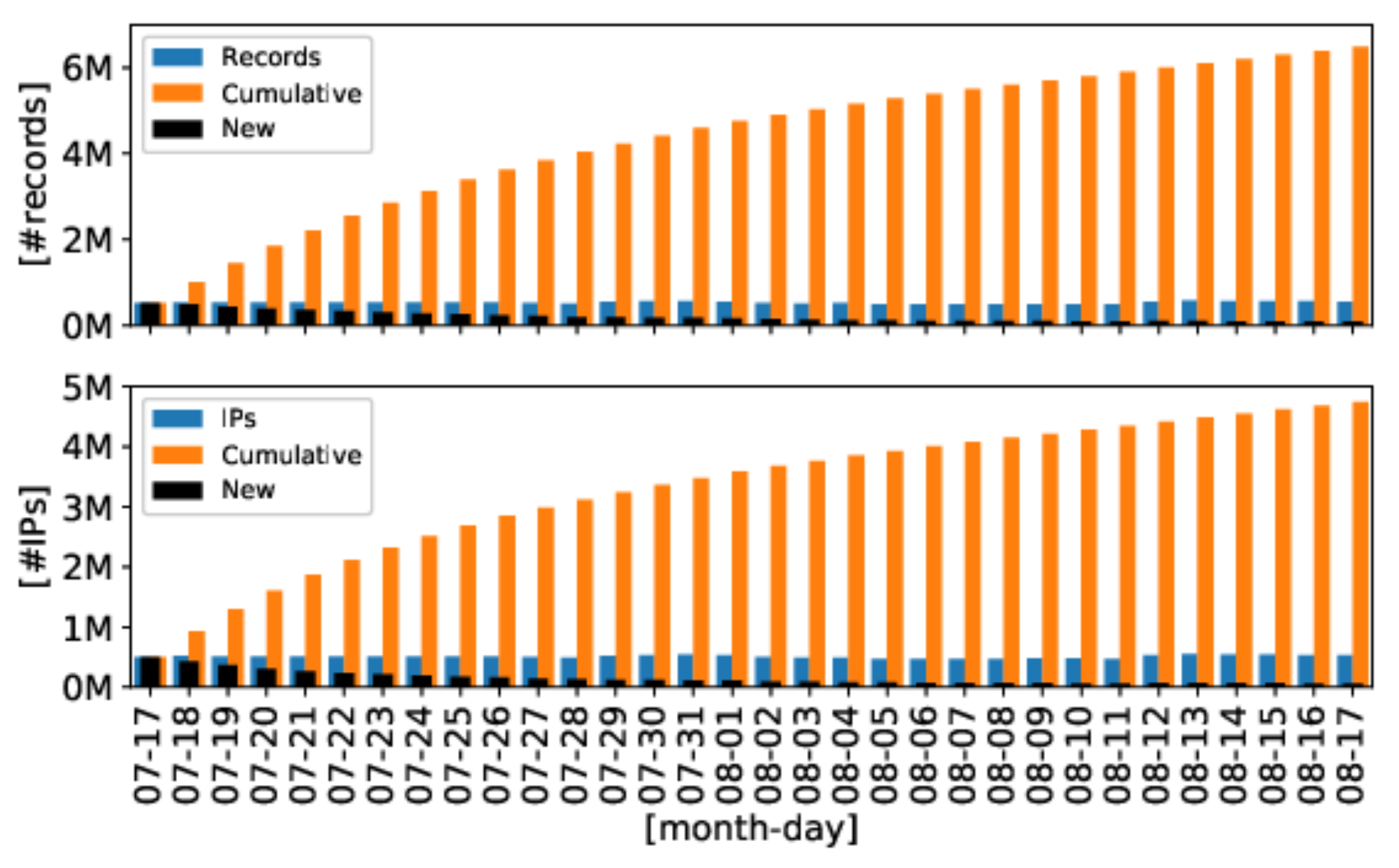}
        \caption{Cumulative number of records (up) and IP addresses (down) related to MikroTik devices, over time.}
	\label{fig:shodan_overall}
\end{figure}

Our observations regarding the number of records and the number of IP
address per day are presented in~\autoref{fig:shodan_overall}. After
one month we had observed 4,742,944 distinct IP addresses and
6,484,420 distinct records related to MikroTik devices. We observed
that both the number of records and the number of IP addresses
scanned/found (blue bars) remained similar from day to day, at
$\thicksim$500k per day (median 535,260 and 510,173, respectively).

In \autoref{fig:shodan_overall}, we also depict the cumulative number
of distinct records (graph up) and IP addresses (graph down); and the
number of new records found in a day (relative to the cumulative)
(black bars). Contrary to what we expected, the number of new records
and IP addresses (relative to the cumulative) is never zero. We
believe that the explanation is probably that we did not ‘flush’ our
observations. After an interval of two weeks, both the number of new
records and the number of new IP addresses seem to stabilise (as
Shodan’s owner told us to expect). We believe that the numbers
represent the potential turnover of MikroTik devices.

In \autoref{tab:ports_shodan}, , we list the 10 ports most commonly
associated with MikroTik devices. Our first observation is that HTTP
(line 3, 6, 8 and 9) runs on various ports, possibly to prevent
malicious access. MikroTik routers are usually managed using the
HTTP(S) protocol. It is also interesting to note that most of the
identified ports are listed as services that run by default on
MikroTik devices (i.e. FTP–21, Telnet–23, SSH–22, HTTP–80, HTTPS–443,
Bandwidth-test–2000, WinBox–8291, API–8728 and API-SSL–
8729)~\cite{mikrotik2019misc}.  However, one of the top ports, 1723,
is not open by default. The port must therefore have been opened
intentionally by the operators or by hackers that got access to the
relevant devices.

Another notable feature of the data in Table I is that the data
gathered by Shodan does not include any entries relating to port 8291
(WinBox), 8728 (API) or 8729 (API-SSL). The reason is that Shodan does
not scan for those ports. The implication is that the number of
devices that we found is potentially smaller than the actual number of
MikroTik devices. However, according to Baines
\cite{baines2019misc}, there are at least 565,648 devices (IP
addresses) running port 8291 (WinBox), which is mainly used by
MikroTik devices. Comparing Baines' findings, available
at~\cite{baines2019misc2}, with our one-month analysis, we observed
more than 80\% overlap (80.17\%). The implication is that, although
Shodan does not cover port 8291, the majority of MikroTik devices run
multiple ports and Shodan is therefore able to detect them.

\begin{table}[h!]
	\centering
	\caption{Top 10 service ports open on MikroTik devices.}
	\label{tab:ports_shodan}
	\begin{tabular}{|c|l|l|c|c|}
	\toprule
	{} &  Port \# & Description & \# Records & \% \\
	\midrule
	1  &         2000 &Bandwidth-test &       3,769,843 &          58.1\% \\
	2  &         1723 &PPTP           &       1,265,191 &          19.5\% \\
	3  &           80 &HTTP           &         410,289 &           6.3\% \\
	4  &           21 &FTP            &         311,952 &           4.8\% \\
	5  &           23 &Telnet         &         164,330 &           2.5\% \\
	6  &         8080 &HTTP           &        139,277 &           2.1\% \\
	7  &          161 &SNMP           &          91,453 &           1.4\% \\
	8  &         8888 &HTTP           &          41,233 &           0.6\% \\
	9  &           81 &HTTP           &          36,292 &           0.6\% \\
	10 &           22 &SSH            &         28,705 &           0.4\% \\
	\bottomrule
	\end{tabular}
\end{table}

Finally, in \autoref{tab:shodan_ips}, we list the 10 countries with
the most IP addresses relating to MikroTik devices. This information
is important for deciding in which countries our honeypot sensors
(described in the next section) should be placed. The ten listed
countries represent 64.88\% of all MikroTik devices that we found.
Note also that 8 out of the 10 countries are considered ‘emerging
economies’ (*). The explanation for the prominence of such countries
may be that MikroTik devices are known as low-cost routers. They are
therefore particularly attractive to ‘emerging economies’ that have a
strong incentive to economise when investing in their network
infrastructures. That hypothesis is supported by the fact that the top
country is Brazil and, according to \cite{certbr2019misc}, half of
the core routers in the largest Brazilian Internet Exchange are
MikroTik devices.

\begin{table}[h!]
	\centering
	\caption{Top 10 countries with IP addresses related to MikroTik devices.}
	\label{tab:shodan_ips}
	\begin{tabular}{|c|l|c|c|}
	\toprule
	{} & Country & \# IP add. &      \% \\
	\midrule
	1  &     BRA* &   759,770 &  16.0\% \\
	2  &     CHN* &   715,325 &  15.1\% \\
	3  &     USA &   272,470 &   5.7\% \\
	4  &     RUS* &   260,553 &   5.5\% \\
	5  &     IDN* &   239,598 &   5.1\% \\
	6  &     ITA &   207,229 &   4.4\% \\
	7  &     IRN* &   197,756 &   4.2\% \\
	8  &     IND* &   153,757 &   3.2\% \\
	9  &     THA* &   137,036 &   2.9\% \\
	10 &     ZAF* &   134,124 &   2.8\% \\
	\bottomrule
	\end{tabular}
\end{table}

\textbf{The takeaway message.} In this section, we observed that the
number of MikroTik devices operating around the world is very high.
They are located largely in emerging economies, and some are used in
the core network infrastructures of such countries.

That observation underscores the importance of the study described in
this paper. In addition to the default service ports open on MikroTik
devices, a realistic MikroTik honeypot should feature the 10 ports
most commonly open, as detailed in \autoref{tab:ports_shodan}.

\section{System design}
\label{sec:honeypotdesign}

In this section, we describe the design of our honeypot, consider
pertinent deployment aspects and discuss the legal considerations.

\subsection{Honeypot design}

\begin{figure}[h!]
    \centering 
    \includegraphics[width=0.7\linewidth]{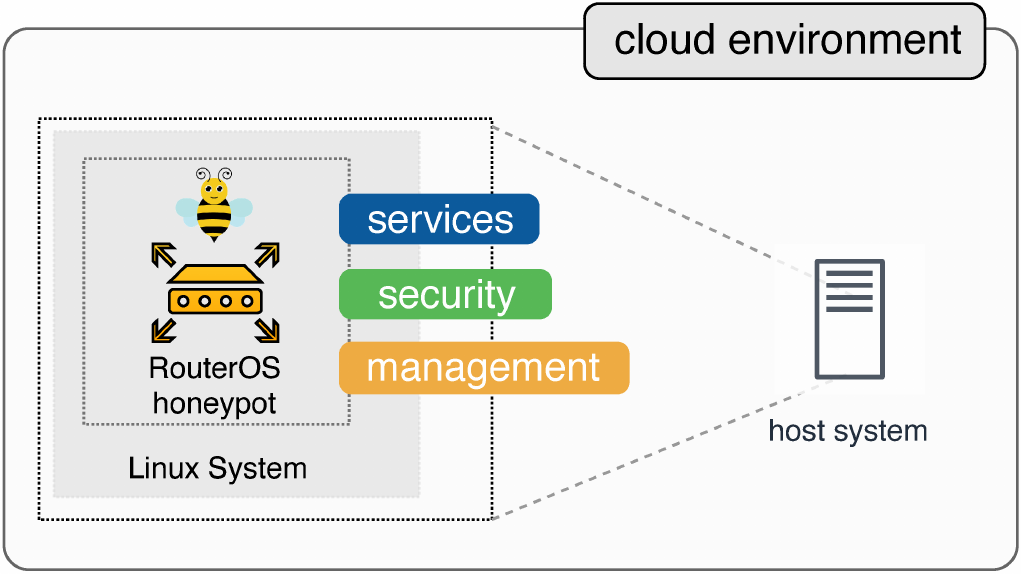}
    \caption{MikroTik honeypot design.}
    \label{fig:honeypot}
\end{figure}

In the interests of reproducibility and deployability, we designed the
honeypot using a system based on paravirtualisation. As depicted in
\autoref{fig:honeypot}, the design features a host system based on
Linux that is responsible for running RouterOS\@. We opted to use
version 6.39.3 of RouterOS, since it has not been patched for most of
the critical vulnerabilities discovered in recent years. The
communication between the host system and RouterOS is performed by a
software API developed using three building blocks described in turn
below.

\noindent\textbf{Services:} the RouterOS system was configured to
provide a set of services to external users. Those services were
deployed in the honeypot and made accessible via the host system. The
host access policy was accordingly updated to enable connection to
those services directly. Although RouterOS enables a set of services
by default, we have customised the system configuration to make the
device more realistic, in line with the findings discussed in
\autoref{sec:landscape}. \autoref{table:mikrotik_default_ports} shows
the list of services enabled in our honeypots.  The services marked
with an asterisk are not active in the default configuration, but were
added in our customisation.

\vspace{2.2cm}
\noindent\textbf{Security:} the security module is responsible for
controlling access to the RouterOS and managing the connections to the
active services. An emulated virtual private network was created by
running the virtual machine in a host-only network to separate the
traffic to the server from that to the honeypot. That ensured that the
honeypot could not gain access to the rest of the server. The honeypot
is a high-interaction honeypot. The advantage of a high-interaction
honeypot is that the chances of receiving and detecting attacks are
higher than with a low-interaction honeypot. The disadvantage is that
more damage could be done to the device and proper security measures
should therefore be taken to ensure that the router can be reset
easily and that the bandwidth is restricted to significantly limit the
damage that attackers can do. For this reason, rate limiting was used
to limit the total volume of traffic to the testbed to 1 Mbps.
Furthermore, the router was restored to its original state every day
at midnight to prevent abuse. 

\noindent\textbf{Management:} our system was designed to collect all
inbound and outbound traffic to and from the honeypot. We collected
the full packets using the TCPDUMP capture tool for each running
sensor and stored them on a centralised server. In addition to the
network traffic, we collected the system events (logs) available in
the RouterOS running in the honeypot. As described by Mazdadi
\textit{et al.} \cite{ROUTEROSFORENSICS:IJCSIS:2017}, it is possible
to use the API interface provided by RouterOS to actively retrieve
relevant system information, such as DHCP leases, configuration
changes, uploaded files, BGP data and more. The tool described in that
paper was not available, so a custom script was written to imitate its
functionality \cite{mikrotik2019exposed}. Our tool was configured to
collect the system events every 5 minutes via the RouterOS API-SSL
service on port 8729.

\subsection{Honeypots deployment}

A total of six honeypots were created in different regions of servers
on the Google Cloud Platform. The locations of the honeypots were
chosen based on the most common regions for MikroTik devices, as per
Table~\ref{tab:shodan_ips}. Another factor influencing location was
that we wanted the honeypots to cover most of the continents. With
that in mind, the regions we settled on were Australia, Brazil, China
(Hong Kong), India, the Netherlands and the United States of America.

A central ‘collector’ computer was used to collect the traffic from
the six honeypots. This computer implemented routines to collect,
filter and parse all the new logs generated by TCPDUMP capture tool as
well as the system events from the honeypot. The information gathering
process was repeated every hour, so even if the honeypot were
compromised and we lost access, we would still have the logs for every
complete hour prior to access loss. It is important to note that, in
that case, our honeypot would be restored using our clean system
snapshot.

\begin{table}[h!]
    \centering
    \caption{RouterOS services enabled in our honeypots.}
    \label{table:mikrotik_default_ports}
    \begin{tabular}{|c|c|l| } 
    \toprule
    \# &Port \# & Service \\ 
    \midrule
    1&2000 & Bandwidth-test\\
    2&1723 & PPTP*\\  
    3&80 & HTTP \\
    4&21 & FTP \\ 
    5&23 & Telnet \\ 
    6&8080 & HTTP-Proxy* \\
    7&22 & SSH \\ 
    8&139 & SMB* \\ 
    9&445 & SMB*\\   
    10&8291 & WinBox \\ 
    11&8728 & API \\ 
    12&8729 & API-SSL \\ 
    \bottomrule
    \end{tabular}
\end{table}

\subsection{Legal considerations}

It is important to consider the applicable legal requirements when
designing a honeypot. According to EU law, “a duty to act positively
to protect others from damage may exist if the actor creates or
controls a dangerous situation” \cite{EUTORTLAW:SPRINGER:2007}. Hence,
a honeypot owner has a responsibility to take appropriate action to
secure the honeypot, since a honeypot operation can be seen as a
potentially dangerous situation insofar as real-world attacks are
attracted.

Research by Sokol and Andrejko
\cite{HONEYPOTSLIABILITY:SPRINGER:2015}  shows that a secure honeypot
meeting the requirements laid down by EU law should have the five
features mentioned below. We used Sokol and Andrejko's findings to
guide the design of the honeypot used in our research:

\begin{itemize}
    \setlength\itemsep{0em}
    \item Firewall: allow connections only to certain ports.
    \item Dynamic connection redirection mechanism: only trusted connections should have access to services outside the honeypot.
    \item Emulated private virtual network: the honeypot should be run in a restricted private network to restrict attackers.
    \item Testbed:  a controlled environment should be used to analyse vulnerabilities in applications.
    \item Control centre: the administrator of the honeypot should monitor connections and quickly respond to incidents.
\end{itemize}

\vspace{0.1cm}
\textbf{Limitations.} As discussed in this section, our honeypots were
placed on a cloud infrastructure (Google Cloud). Some attackers may
avoid well-known IP address ranges, such as those used by cloud
service providers. Some attackers could also avoid our honeypots
because they are not actual routers (not providing last-mile access).

\section{Attack Classification, Findings, and Discussion}
\label{sec:results}

In this section, we first present our methodology for automating the
classification of attacks on MikroTik devices. Then we present our
observations over 120 days of data collection.

\subsection{Attack Classification Methodology}

Our methodology for classifying attacks relies on comparing the
collected data with  manually created signatures. We created two
databases of attacks containing: (1) signatures that cover the
majority of Common Vulnerabilities and Exposures (CVE) involving
MikroTik devices (as listed by \cite{CVELIST}); and (2) signatures of
known attacks that target low-cost routers in general.

For our attack classification, we consider two types of input file
collected by our MikroTik honeypots: (1) packet-based network traces
(\textit{.pcap} files) and (2) logs. While \textit{.pcap} files are
used for classifying CVE-MikroTik-related attacks, logs are used for
classifying generic attacks. The signatures generated from
\textit{.pcap} data were made compatible with the Berkeley Packet
Filter; they are publicly available from the project
website\footnote{See: \url{http://mikrotik-exposed.org/}}.  All the
signatures developed were validated using proof-of-concept
vulnerability traffic by Tenable
Research~\cite{CVE2018-1156:Tenable:2018}, which is the best-known
dataset for validating attacks on MikroTik devices.

\begin{figure}[h]
    \centering
	\includegraphics[width=0.8\linewidth]{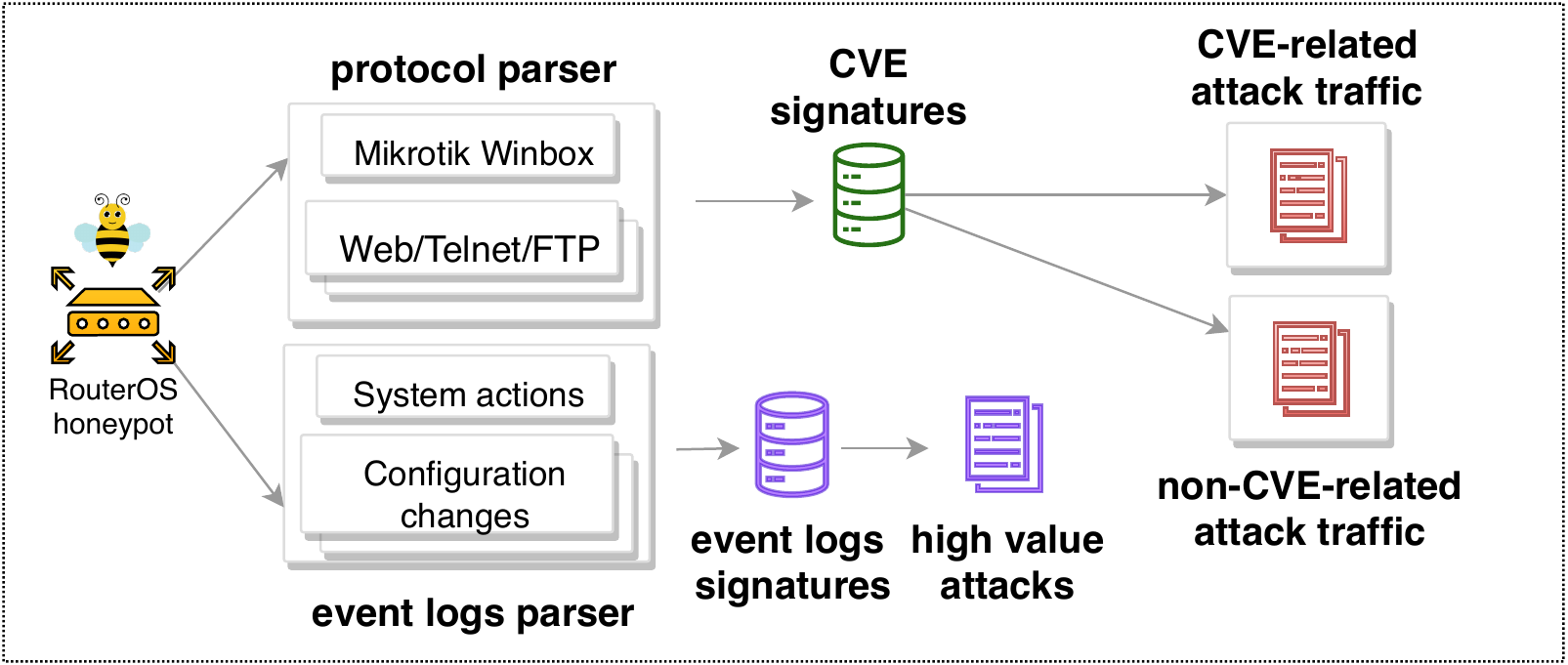}
	\caption{Automated attack classification process.}
	\label{fig:process}
\end{figure}

In order to make full use of the \textit{.pcap} files, we decrypted
any traffic encrypted on the basis of the MikroTik proprietary
protocol. For example, packets based on the WinBox protocol were
decrypted using the tool proposed by Tenable
Research~\cite{winbox:parser-2019-3943:TENABLE:2019} and any
encrypted web traffic was decrypted using the tool proposed by Tenable
Research~\cite{json::parser-2019-3943:TENABLE:2019}.
\autoref{fig:process} depicts our entire methodology.

\textbf{Dataset.} We collected 120 days of data (.pcap and logs), from
25 July 2019 to 20 November 2019, relating to traffic targeting our 6
honeypots (in Australia, Brazil, China, India, Netherlands and United
States of America). In total, we collected more than 44 million
packets (12 million flows) and 3.8 million log records. The authors
will be glad to share the entire dataset on request, for research
purposes.

\textbf{Limitations.} It is important to stress that we did not focus
our classification on the overall backscatter traffic collected by our
honeypot, as most previous researchers in this field have done.
Instead, we analysed a small, critical portion of the entire dataset,
consisting mainly of attack traffic tailored to MikroTik devices.
Another limitation is that we were able to generate signatures for
only half of the known CVEs. Hence, we covered only some of the
attacks tailored to MikroTik devices. Also, more sophisticated attacks
may have been missed.

\subsection{General Traffic Analysis}

\begin{figure}[h]
	\includegraphics[width=\linewidth]{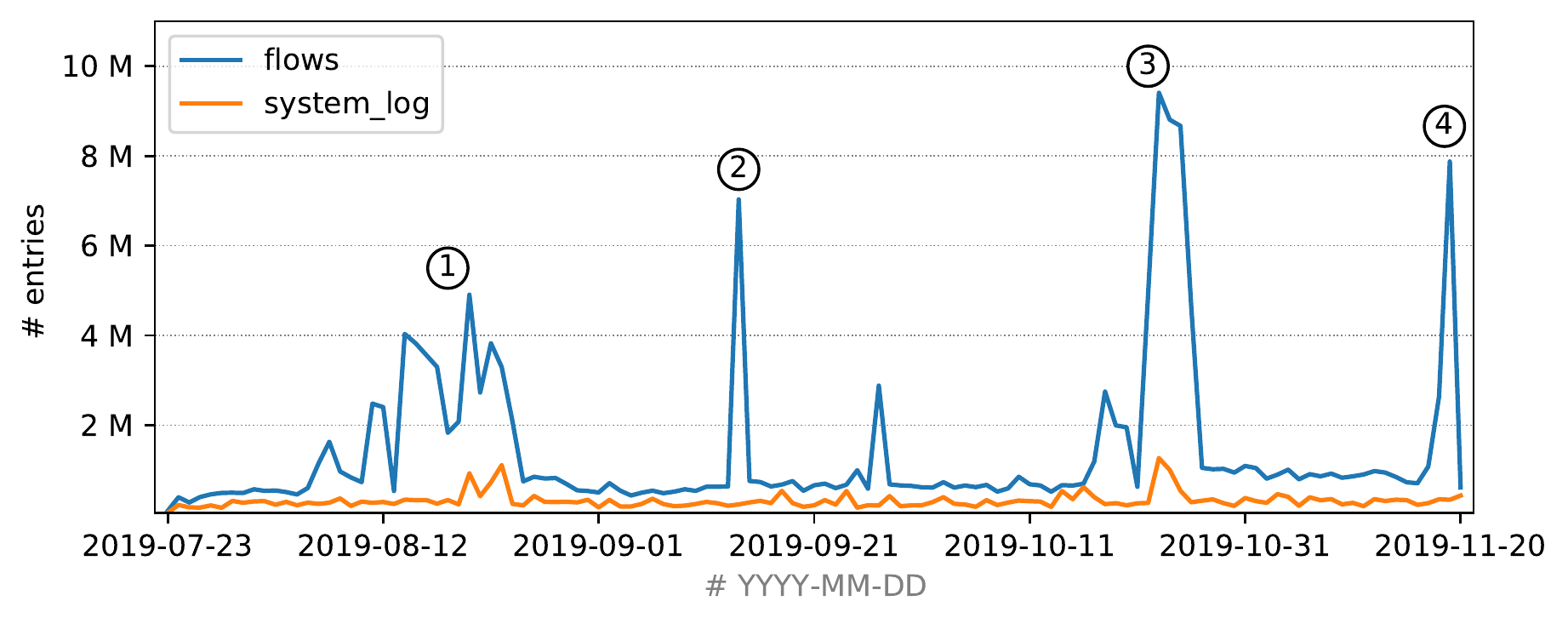}
	\caption{Dataset collected and analyzed using our methodology.}
	\label{fig:dataset}
\end{figure}

\autoref{fig:dataset} shows the timeline of collected data. It
combines information from the \textit{.pcap} files and system log
entries representing hostile activities targeting the MikroTik system.
The number of flows is not constant, therefore the number of system
log events is not constant either. The peaks representing attack
campaigns are considered later in this paper. In the meantime, note
the correlation between the number of flows and the number of log
events in some periods, such as at juncture \ding{174}. The
correlation indicates that the attack campaign in question
successfully generated honeypot system activity.

\textbf{Attack campaigns}. A campaign consists of attackers performing
multiple offensive actions targeting a specific resource or
vulnerability. Such behaviour typically results in outliers in our
dataset, as seen at the numbered junctures in  \autoref{fig:dataset}.
In 120 days of analysed data, 4 campaigns stand out. Campaign
\ding{172} was performed by one unique IP (64.X.X.70) and targeted the
MikroTik management service API (8728/TCP). The IP in question hit all
our honeypots equally. The campaign probably involved a horizontal
scan of the internet to enumerate MikroTik-related services. The next
identified campaign was also conducted from one unique IP address
(183.X.X.171). In this campaign, several TCP ports were targeted.
Notably, all the probes involved in this attack used the same source
port 55453/TCP\@. Since use of the same source port for all flows is not
the default behaviour, we suspect the use of a customised and
automated tool to perform the scans. The same behaviour can be seen in
campaign \ding{174}: only one IP address performed a scan using a static
source port. Campaign \ding{175} involved brute-forcing attacks
targeting the FTP service and was conducted from a single IP address
(185.X.X.34). The identification of such campaigns helps us to
understand the threat landscape. Although we did not find a
correlation between the described campaigns and vulnerability
exploitation, the analysis can provide insight regarding the way
attackers operate.

\textbf{Attack dispersal}: we investigated how the attackers hit our
honeypots. By analysing the IP addresses used and the honeypots
targeted, we can determine the attack dispersal. We found that 80\% of
attackers were associated with a maximum of 2 honeypots. Whereas 67\%
of source IPs hit only one honeypot.

\begin{figure}[h!]
    \centering
    \includegraphics[width=0.8\linewidth]{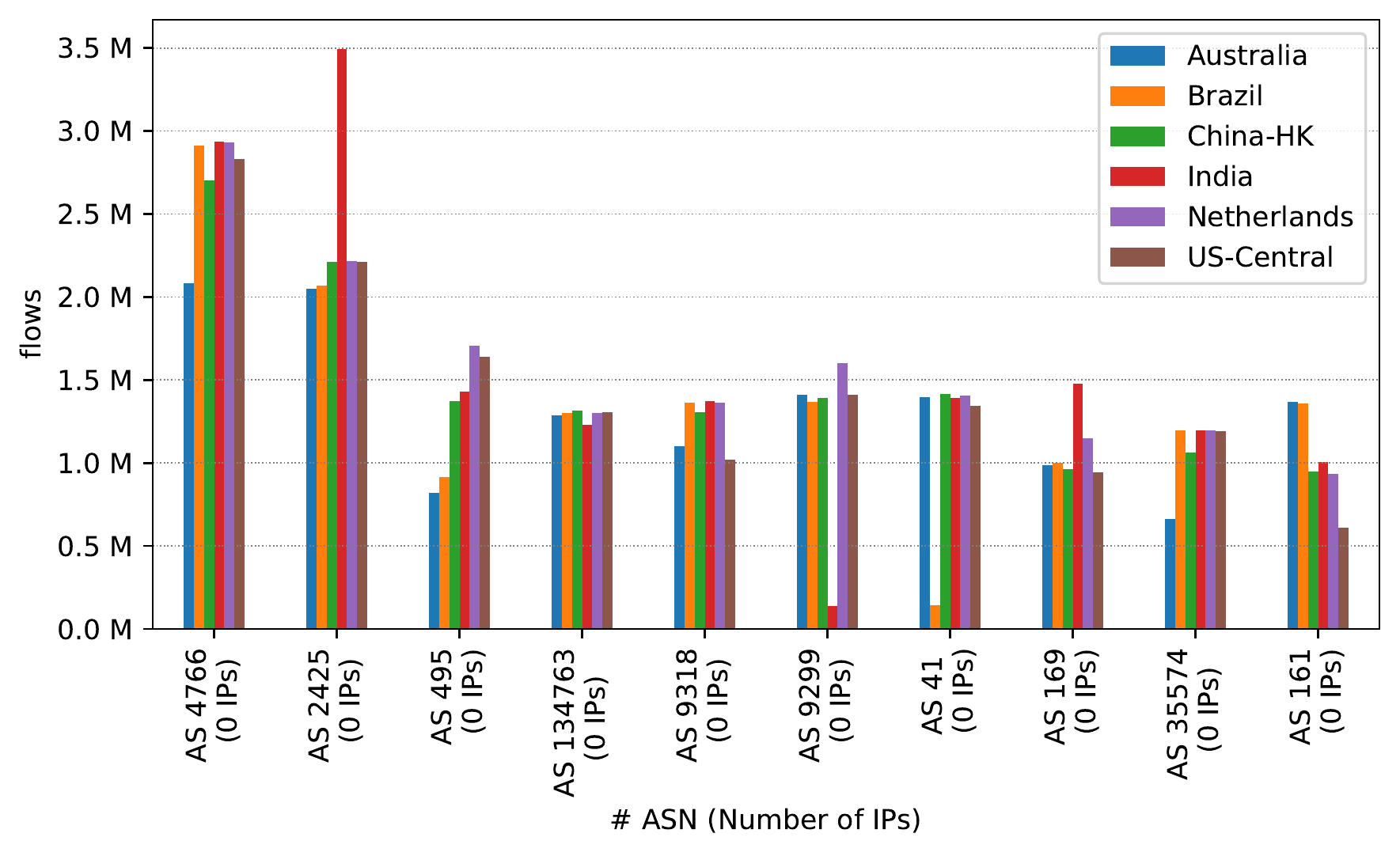}
    \caption{Attack dispersal associated with MikroTik devices, grouped by origin (AS).}
    \label{fig:map_attack_location}
\end{figure}

At the AS level, \autoref{fig:map_attack_location}, the attacks are
fairly dispersed as well.  For example, the most prominent AS (AS
4766: Korea Telecom) is operated by a large ISP in Korea, while the
third most prominent AS (AS134763: Chinanet) belongs to an ISP located
in China. The second most prominent AS is located in the Seychelles
(AS 202425: IP Volume Networks) and seems to be a data centre that
provides suspicious services, perhaps a \textit{bulletproof} service
or some service model that offers attacks tailored to MikroTik. Not
much information is available regarding the AS's business model,
however; the operators accept only Bitcoin payments.

\subsection{MikroTik CVE-Related Attacks}
\label{sec:cve}

We have identified 3,441 attacks related to CVEs of MikroTik devices.
The attacks represent only 0.02\% of the total number of flows
collected. Although the attacks represent a very small percentage of
the total traffic, they were tailored to MikroTik devices. That means
that the attacks would have caused more damage to real devices and
their users than generic attacks.

The most popular attack vector that we observed was ‘directory
transversal’ attack (related to CVE-2018-14847 and CVE-2019-3943).
Such attacks enable attackers to access restricted files and
directories within the router. By investigating the payload of the
attacks, we observed that 98\% of them successfully acquired the
credentials of administrator accounts. The remainder sought to change
the system job scheduled for executing commands. The commands involved
are considered in more detail later.

In the next subsection we focus on the successful login attempts made
after attackers exploited the vulnerabilities referred to (CVE-2018-
14847 and CVE-2019-3943). Although some vulnerabilities were mapped to
signatures, we did not observe any attack that sought to perform
remote code execution. For example, MikroTik RouterOS has two
vulnerabilities that enable attackers to run arbitrary code or
commands in the system (CVE-2018-7445 and CVE-2018-1156). The
vulnerabilities involve the NETBIOS protocol and a specific service
module triggered by a particular request to port 80/TCP.

\subsection{Successful Login}
\label{sec:login}

\begin{figure}[h]
    \centering
    \includegraphics[width=0.8\linewidth]{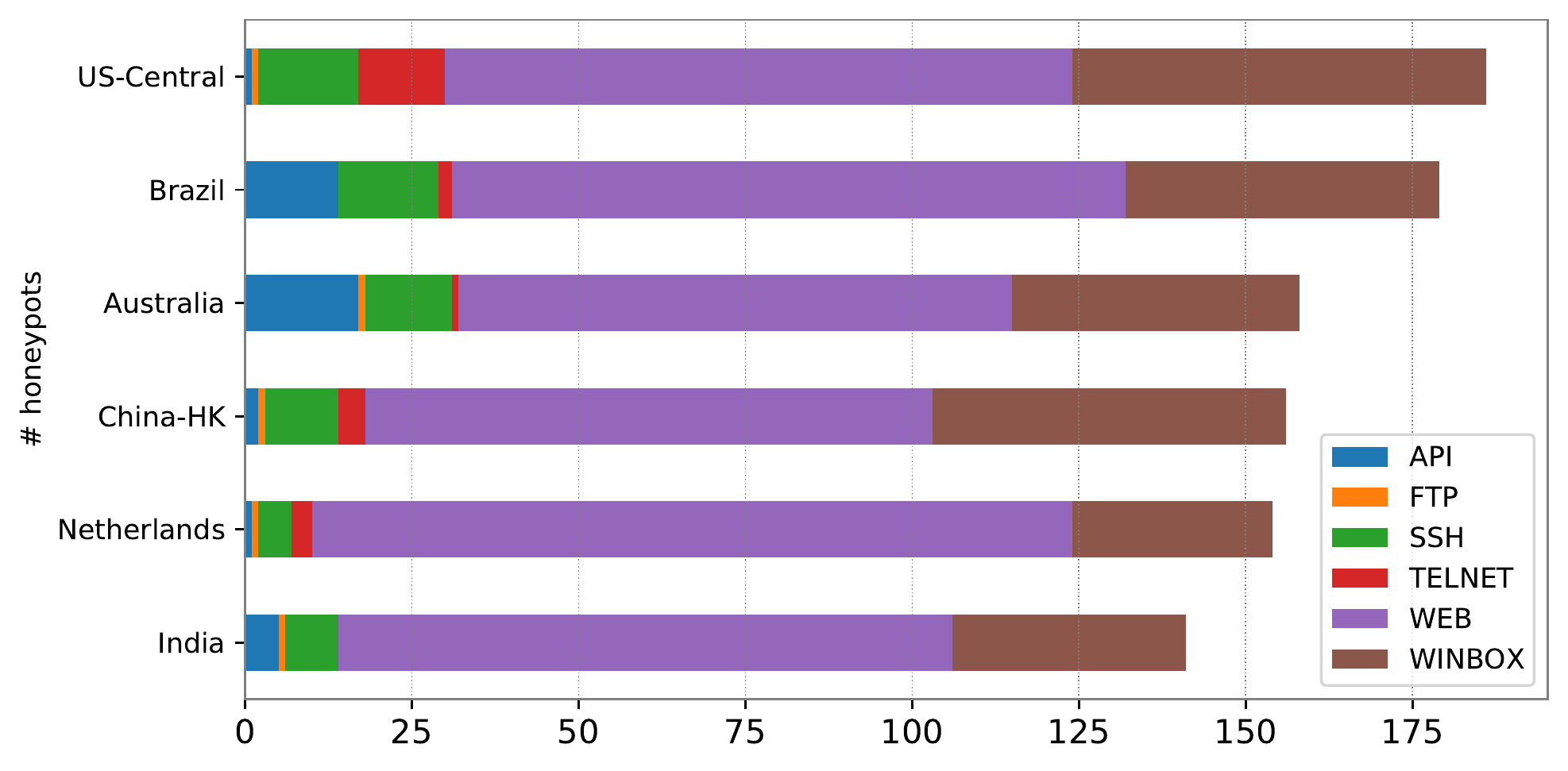}
    \caption{The number of successful login attempts in our honeypots.}
    \label{fig:login}
\end{figure}

Even though we used a strong and not easily guessed password (16
random letters and digits, with capital and non-capital letters), we
observed 983 successful logins, depicted in  \autoref{fig:login}. The
figure shows, for each honeypot, the distribution of services
associated with successful logins (API, FTP, SSH, Web/HTTP and
WinBox). In contrast to the successful logins, we also observed
exhaustive attempts to log in to MikroTik-specific services. However,
we did not map successful brute force attacks that used our specific
password.

The honeypots hosted in the US and Brazil were the ones that were
successfully compromised most often. That is interesting, since Brazil
is the country with the most discoverable MikroTik devices, as pointed
out in \autoref{sec:landscape}.

Logins using the WinBox and HTTP interfaces were preferred by the
attackers attracted to our six honeypots. The most plausible
explanation for that observation is that the two protocols are the
easiest for managing MikroTik devices (via a visual interface).
Targeted attacks usually use automated tools to connect to the system
and establish a channel with the attacker. Most such tools used for
MikroTik devices are based on WinBox or the Web API (HTTP). Attempts
made using the FTP and SSH protocols provide a more restrictive
interface for managing RouterOS services when compared with MikroTik
protocols (e.g. WinBox).

\autoref{fig:login}  also shows that the distribution of attacks is not uniform
across all honeypots. For example, no login attempts were made using
Telnet in India, although the method was used elsewhere. The
heterogeneous distribution of authentication methods suggests multiple
attack vectors, resulting in various automated tools for various
attacker \textit{modi operandi}. The various attack vectors are
considered in the next subsection.

\subsection{Successful Tunnel Creation} 
\label{sec:tunnel}

\begin{figure}[h!]
    \centering
	\includegraphics[width=0.8\linewidth]{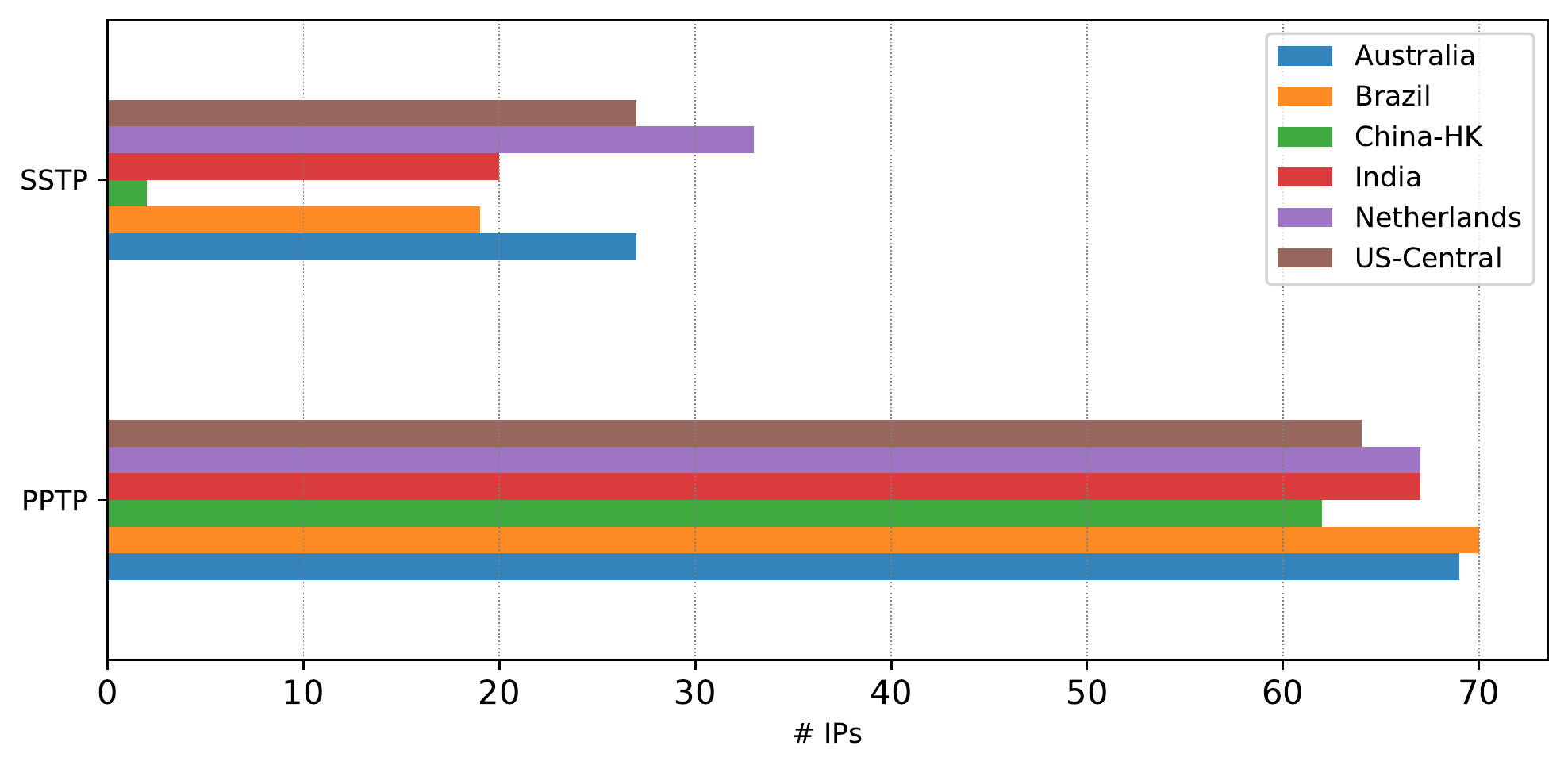}
	\caption{Numbers of established tunnels by honeypot, grouped by protocol (SSTP/PPTP).}
	\label{fig:tunntype}
\end{figure}

In the work described thus far, we observed that successful
login-related attacks (\autoref{sec:login}) occurred after the
exploitation of known CVEs (\autoref{sec:cve}). Following this chain
of attacks, we also observed traffic tunneling-related attacks. The
usage of tunnels is a well-known technique in the context of attacks
against routers. It involves attempting to redirect, intercept or deny
network traffic from/to routers. In our honeypots we observed the most
common tunneling protocols, Point-to-Point Tunneling Protocol (PPTP)
and Simple Service Discovery Protocol (SSDP) have a similar abnormal
\textit{modus operandi}. The attackers first sign into the system via
the WinBox protocol using the correct credentials and then set up a
tunneled connection to the outside end-point for exporting traffic.
We have identified 3,137 successfully established tunnels, 2,171 of
them using PPTP and 966 using SSTP\@ (See
\autoref{fig:tunntype}). PPTP was the most popular. Interestingly the
protocol PPTP is known for being insecure. The protocol is old and
vulnerable. The traffic can easily be decrypted by malicious third
parties in man-in-the-middle attacks. Unlike PPTP, SSTP uses SSL 3.0,
providing a strong encryption scheme.

\autoref{fig:tunntype} shows the distribution of tunnels across the
honeypots, grouped by protocol type. In China, for example, the number
of attacks using SSTP is very low compared with other regions. Since
all honeypots have both services available, is up to the attacker to
choose the protocol. We presume that PPTP is more popular due to easy
setup (it does not require SSL configuration). In the next subsection
we discuss the attribution of such attacks in detail.

\subsubsection*{Attributing Successful Tunnel Creation}

\begin{figure}
        \centering
	\includegraphics[width=0.8\linewidth]{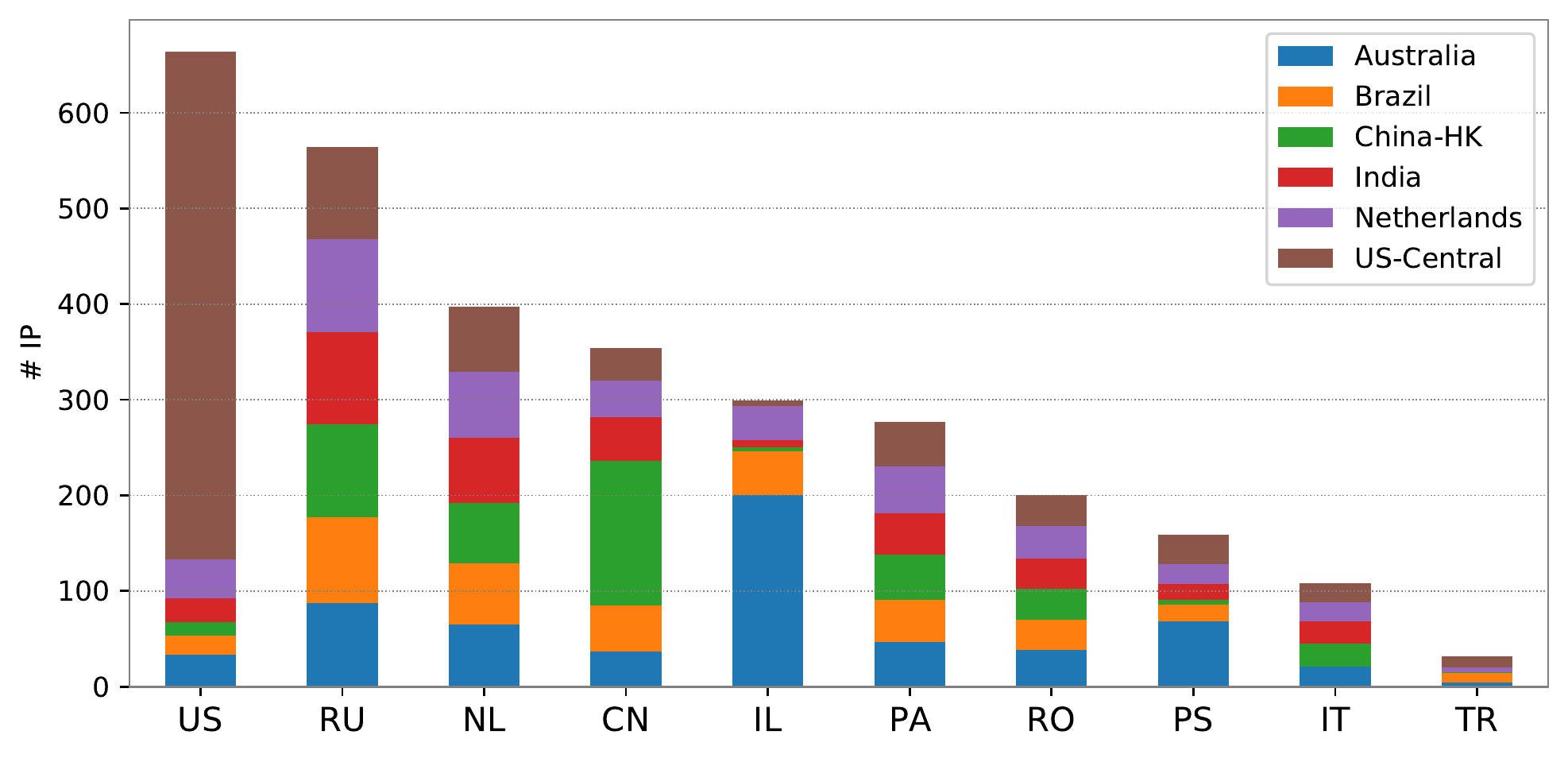}
	\caption{Tunnel endpoint country code.}
	\label{fig:tunnel}
\end{figure}

In \autoref{fig:tunnel}, we depict the distribution of IP addresses
relative to the tunnel endpoint country code. We observe that IP
addresses located in the United States, Russia and the Netherlands
represent more than 50\% of all IP address exploiting traffic tunnels.
An interesting case is presented in the US\@. Out of 664 tunnels
endpoints located in US, 80\% (531) of them were used only in attacks
performed in the US honeypot.

\begin{figure}
        \centering
	\includegraphics[width=0.8\linewidth]{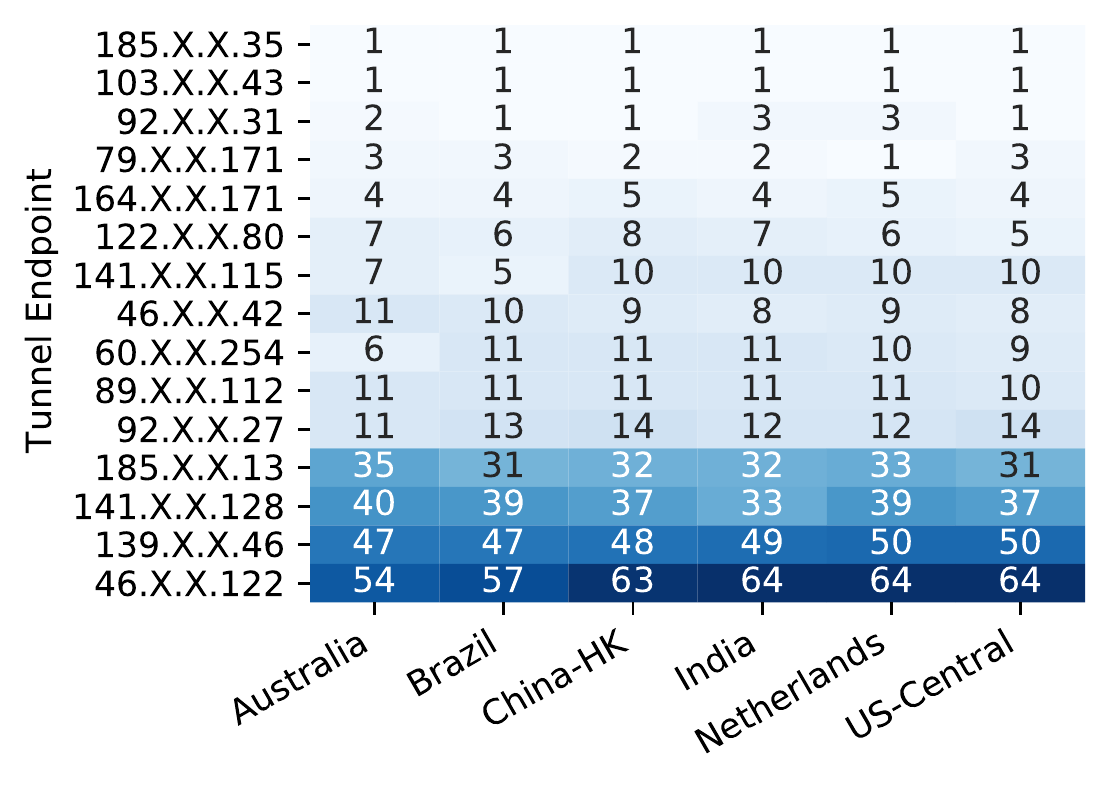}
	\caption{Endpoint-honeypot correlations.}
	\label{fig:heatmap}
\end{figure}

With a view to shedding light on correlations with endpoints, in
\autoref{fig:heatmap}, we present data on the number of times the top
15 IP address were used as tunnel endpoints. Surprisingly, 5 IP
addresses are responsible for almost 50\% of the total number of
tunnels. The IP address 139.X.X.46 was used as an endpoint in all the
honeypots and it was used in 116 days of analysis (out of 120 days).
It is important to remember that our honeypot was rebuilt/re-initiated
every day, forcing an attacker to compromise the system again to
establish a tunnel.

Unlike the sources of targeted attacks that exploit vulnerabilities,
the tunnel endpoints are concentrated in a relatively small set of
IPs. IPs located in certain countries (\autoref{fig:tunnel}) and
certain endpoints (\autoref{fig:heatmap}) are more likely to be used
in all the honeypots.

\subsection{Brute-Force Attacks and Mirai Botnet}

As well as attacks tailored to MikroTik, we have investigated
brute-force attacks, since MikroTik devices do not have a protection
mechanism against such attacks. In the traffic that we collected in
our honeypots, the SSH service (22/TCP) was most targeted by
brute-force attacks, with 2,287,479 attempts. The Telnet service
(23/TCP) was the second most targeted service, with 1,848,845
attempts. By analysing the IPs that were used for scanning on SSH and
Telnet, we were able to observe that there is an overlap of 39\%. That
indicates that 39\% of attacks were targeting both services.

\begin{figure}[h]
        \centering
	\includegraphics[width=0.8\linewidth]{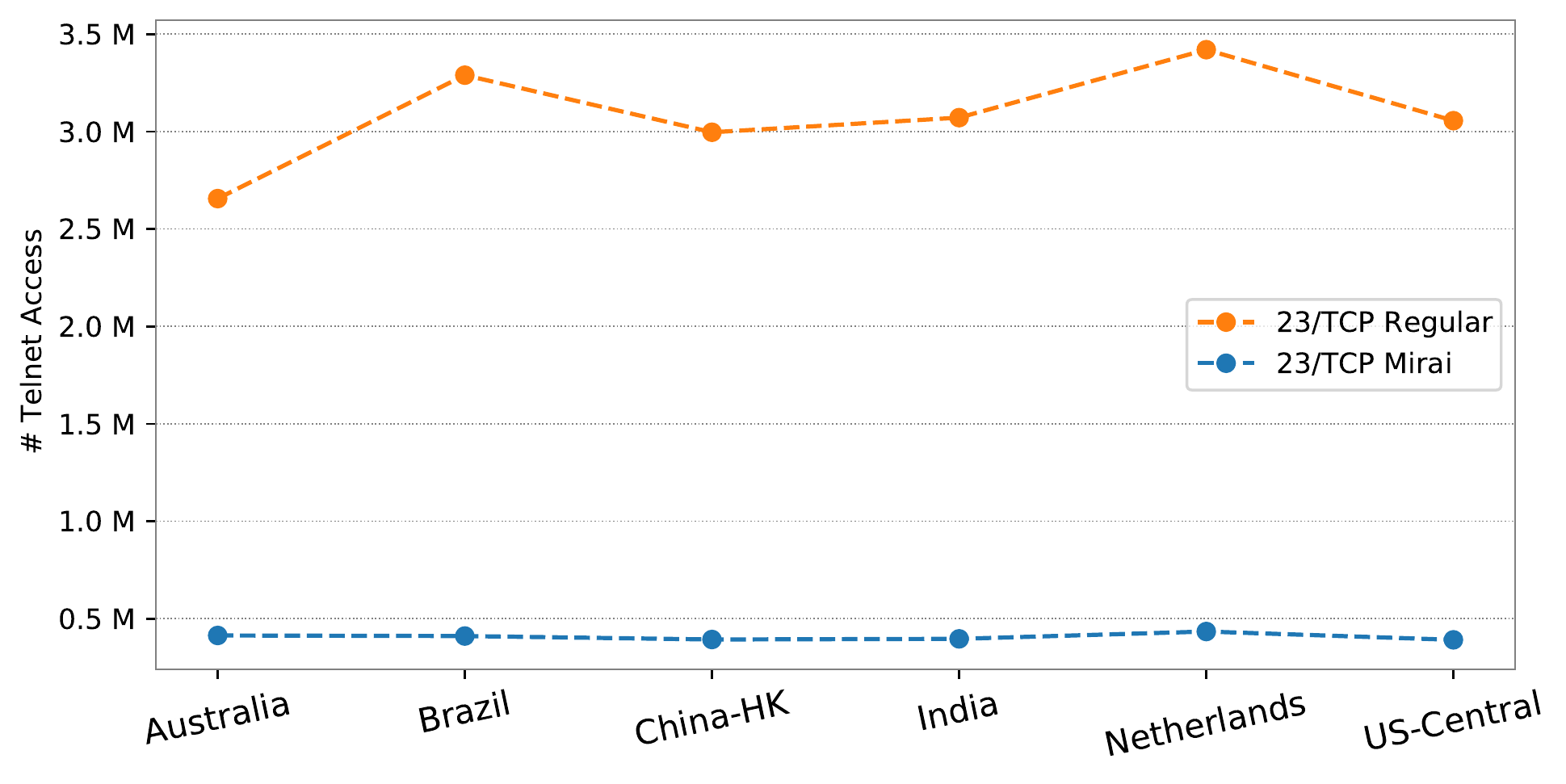}
	\caption{Brute-force attacks targeting 23/TCP (Telnet).}
	\label{fig:telnet}
\end{figure}

A large subset (13\%) of brute-force attacks on Telnet used a known
signature of the Mirai botnet~\cite{IMPROVINGIOTBOTNET:SENSORS:2018}.
\autoref{fig:telnet} shows the number of Telnet attempts by honeypot,
grouped into regular brute-force attacks and attacks using the Mirai
signature. While there are some variations in the number of regular
attacks, the number of Mirai attacks is quite stable (bottom line).
That is interesting, insofar as it reveals the level of internet noise
caused by compromised IoT devices performing scans.

Mirai is an IoT malware that tries to compromise devices using
brute-force attacks and turn them into bots. We observed 101,443
unique IP addresses that were infected by Mirai. Figure 11 shows the
distribution of IPs performing Mirai brute-force attacks. IP addresses
in China account for more than 25\% of attacks. The majority of those
IPs belong to Transit providers in China, such as China Telecom,
Chunghwa and China Unicom. Since Mirai is tailored to IoT devices, the
mapping reveals that many IoT devices operated by end-users (Transit
providers) were infected and performing brute-force scans globally.

\begin{figure}[h]
        \centering
	\includegraphics[width=0.8\linewidth]{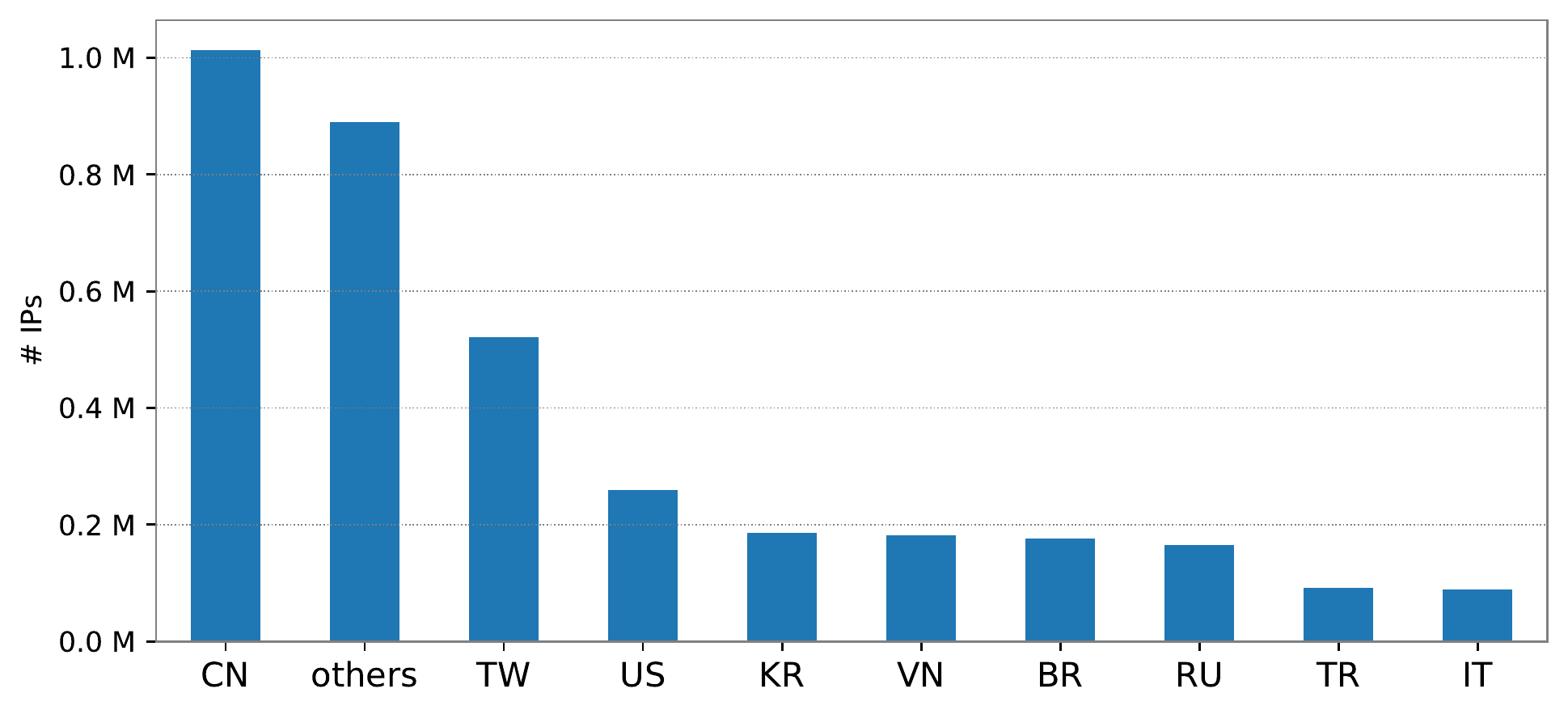}
	\caption{Number of IPs by country code performing brute-forcing attacks with the Mirai fingerprint.}
	\label{fig:mirai}
\end{figure}

\subsection[Log-based Analysis]{Log-Based Analysis for Validating our Observations} 

With the aim of validating the attack data presented in the previous
sections (based on packet-based measurement) and possibly identifying
other attacks, in this section we analyse the content of the system
event logs from our honeypots. \autoref{table:events1} summarises some
of our findings.

\begin{table}[h!]	
	\centering
	\caption{System actions performed by attackers.}
    \scalebox{0.85}{
    \begin{tabular}{llr}
    \toprule
    {} &                                        message &  count \\
    \midrule
    1  &                     new script scheduled by admin &    154 \\
    2  &             PPTP server settings changed by admin &     63 \\
    3  &             L2TP server settings changed by admin &     50 \\
    4  &             SSTP server settings changed by admin &     38 \\
    5  &                              DNS changed by admin &     25 \\
    6  &                               DHCP client changed &     21 \\
    7  &                      PPTP server settings changed &     19 \\
    8  &                      SSTP server settings changed &     19 \\
    9  &                      L2TP server settings changed &     19 \\
    10 & PPP profile \textless default-encryption\textgreater changed by admin &     19 \\
    \bottomrule
    \end{tabular}
    }
	\label{table:events1}
\end{table}

The first observation is that the log files are unstructured, which
makes the pattern identification process challenging. Nevertheless,
from \autoref{table:events1} we observe that lines 2, 3, 4, 7, 8 and 9
emphasise changes in the tunneling (discussed in
\autoref{sec:tunnel}). The most common event in the system log (Line
1), however, involves an attacker adding a script (crontab) to allow
further control of the MikroTik device. The use of scheduled scripts
is considered in more detail below. Note that all the events listed in
\autoref{table:events1} happened after successful logins (described in
\autoref{sec:login}), which were only possible because attackers
exploited known MikroTik CVEs (described in \autoref{sec:cve}).

\textbf{Script scheduled}: By analysing the system logs, we identified
154 events where a script was successfully scheduled on the system. To
complement the analysis of such events, we also inspected the network
traffic for keywords that might reveal the script scheduler. We found
30 new events that were not mapped in the system log. The explanation
for that observation is that other scheduled scripts modified the
device and made outside connections unavailable. Thus, we were unable
to retrieve the log entries for those attacks.

A total of 184 attacks using the script scheduler were discovered. Of
those, 175 used the fetch tool to download scripts and run them in the
system. The scripts were downloaded from nine different websites,
listed in \autoref{table:scripts_sites}.

\begin{table}[h!]	
	\centering
	\caption{Top sites hosting downloaded scripts.}
    \begin{tabular}{llr}
    \toprule
    {} &                    site &  count \\
    \midrule
    1  &            7standby.com & 38 \\
    2  &            phonemus.net & 37 \\
    3  &            hitsmoby.net &  36 \\
    4  &            1awesome.net &  32 \\
    5  &            takebad1.com &  32 \\
    6  &           up0.bit:31415 &  27 \\
    \bottomrule
    \end{tabular}
	\label{table:scripts_sites}
\end{table}

None of the URLs used for downloading were still active when we
performed our analysis. Evidently the URLs were created to propagate
malicious software and were not third-party compromised systems.

The scheduled scripts contain multiple system actions that are
executed in the attacks, besides the system actions instigated by the
scripts downloaded using the fetch tool.
\autoref{table:script_commands}  provides an overview of the most
commonly executed system actions. The actions usually include enabling
the HTTP proxy, changing the DNS and NTP servers, removing existing
crontab scripts and closing API access to the router. Other actions
include disabling FTP, SOCKS and the built-in packet sniffer. Finally,
some attackers also created a new user to maintain easy access to the
router.

\begin{table}[h!]
	\centering
	\caption{Top system actions in scheduled scripts.}
    \begin{tabular}{llr}
    \toprule
    {} &                message &  count \\
    \midrule
    1  &            /tool fetch & 240 \\
    2  &      /system scheduler &  81 \\
    3  &           /file remove &  40 \\
    4  &              /ip socks &  33 \\
    5  &           /ip firewall &  30 \\
    6  &              /user set &  22 \\
    7  &           /user remove &  20 \\
    8  &            /ip service &  20 \\
    9  &              /ip proxy &  20 \\
    10 &                /ip dns &  20 \\
    \bottomrule
    \end{tabular}
    \label{table:script_commands}
\end{table}

\textbf{Cryptocurrency mining}: we identified 9 attacks that injected
scripts related to cryptocurrency mining. All 9 attacks were performed
from one IP address in China and involved the injection of HTML code
snippets directly into the router. The code snippets were then used to
inject the Coinhive cryptocurrency mining software into the webpages
visited. The commands used a common modus operandi: first the attacker
enables an HTTP proxy on the router. Then the attacker changes the
firewall and NAT settings to redirect all traffic through the proxy.
The script is then used to inject the miner into the sites. Finally,
the attacker closes the FTP and SOCKS services and the built-in packet
sniffer.

\textbf{DNS redirection}: we found several attacks where an attempt
was made to change the router's DNS server (\textit{DNS Changer}
attack). The aim of such an attack is to redirect users to phishing
websites and malvertising~\cite{meng2013dns}. We identified 5 IP
addresses that were used for such attacks. Most of them were already
inactive when this analysis took place. However, one IP was still
active (94.X.X.254).  The IP address in question belongs to the
OpenNIC~\cite{opennic} initiative.  OpenNIC offers an alternative to
the DNS root servers currently in use. We could not see a clear motive
for the attack, other than overwriting an existing DNS domain.
Furthermore, the OpenNIC IP address is classified as an open resolver
and could be used in amplification attacks.

\subsection{Overall Discussion}

Our classification methodology has proven valuable for identifying and
classifying attacks tailored to MikroTik devices. In the same way that
we used signatures to classify attacks identified in offline data,
signatures could be used for active blocking of such attacks. We
observed that attacks targeting MikroTik devices follow a chain of
problems, which always begin with failure to update MikroTik
RouterOS\@. Since 2011, MikroTik has released a total of 138 stable
versions of RouterOS\@.

All the current 22 CVEs could be resolved simply by updating the
RouterOS version. That would, for example, mitigate the revelation of
administrator credentials, thus mitigating successful logins and hence
the creation of traffic tunnels and so on.

Despite the simplicity of the solution, there are thousands of
MikroTik devices with old RouterOS versions. We estimate that more
than two million MikroTik devices are still vulnerable to the majority
of CVEs. We hope that the findings presented in this paper promote an
improvement in the security status of the MikroTik device park.

\section{Conclusions and Future Work}
\label{sec:conclusion}

In this paper, we have examined attacks targeting MikroTik devices. To
investigate such attacks, we developed an automated classification
method based on network signatures and system log events. To improve
the quality of the results we (1) investigated the global landscape of
MikroTik devices using 120 days of data from an internet scanning
service (\url{shodan.io}); (2) designed a realistic honeypot that
mimics the characteristics of a MikroTik RouterOS discoverable over
the internet.

To validate our approach, we deployed 6 honeypots and placed them in
countries where we had established that such devices are popular. By
using the honeypots for 120 days, we collected more than 44 million
packets originating from 399,235 unique IP addresses. Using this
dataset, we evaluated our proposed classification methodology and
identified targeted attacks. In total, we identified 3,441 attempts to
exploit well-known vulnerabilities and about 200,000 IPs performing
brute-force attacks on our honeypots.

The majority of vulnerability exploitations involve attempting to
retrieve the credentials and then use them to manage the system
remotely. We have observed various actions that attackers perform on
the system, including changes to the packet filter configuration and
system event logs. For the authors, one of the most interesting
findings is that attackers often seek to establish IP tunneling on
compromised devices. They use the tunnels to redirect traffic and
secretly monitor/inspect/manipulate the data. We have identified more
than 3,000 successfully established tunnels on the honeypots.

The protocol most commonly used for attacks detected in our honeypots
was PPTP (173/TCP), a service that is not activated by default on
MikroTik devices. Surprisingly, as revealed by our investigation using
data from \url{shodan.io}, port 173/TCP is the second most commonly
used port on MikroTik devices worldwide (984,349 devices). That
suggests that either the administrators in question are manually
activating the PPTP service, or those devices have been compromised.
That question requires further investigation.

In the research described in this paper, only attacks on low-cost
MikroTik routers were analysed. In future research, honeypots
simulating other brands of low-cost router could be used to discover
if there are differences in the characteristics of attacks aimed at
different vendors' products. In the context of any future
investigation, consideration should be given to placing honeypots
primarily on an access network where a different class of attack might
be mapped.

\section*{Acknowledgments}

This research was supported partly by EC H2020 GA 830927 (CONCORDIA
project) and by SIDN Fund 174058 (DDoSDB project). We are particularly
grateful to John Matherly of Shodan.io, who promptly provided us with
a unique dataset.

\bibliographystyle{unsrt}
\bibliography{wileyNJD-AMA.bbl}%

\clearpage
\end{document}